\documentstyle[referee]{mn}

\input{psfig.sty}

\def \be{\begin{equation}} 
\def \ee{\end{equation}} 
\def \bea{\begin{eqnarray}} 
\def \eea{\end{eqnarray}}

\begin{document}

\title[Hyperbolic character of the angular moment equations]
{Hyperbolic character of the angular moment equations of radiative
transfer and numerical methods.}

\author[J.A. Pons, J. M$^{\underline{\mbox{a}}}$ Ib\'a\~nez
and J. A. Miralles]{J. A. Pons$^{1,2}$, J. M$^{\underline{\mbox{a}}}$
Ib\'a\~nez$^1$ and J. A. Miralles$^1$ \\
$^1$ Departament d'Astronomia i Astrof\'{\i}sica, 
Universitat de Val\`encia, 
46100 Burjassot, (Val\`encia), Spain \\
$^2$Department of Physics \& Astronomy, SUNY at Stony Brook,
Stony Brook, New York 11794-3800 }

\maketitle

\begin{abstract}

We study the mathematical character of the angular moment equations 
of radiative transfer in
spherical symmetry and conclude that the system is hyperbolic
for general forms of the closure relation found in the literature.
Hyperbolicity and causality
preservation lead to mathematical conditions allowing to 
establish a useful characterization of the closure relations.
We apply numerical methods specifically designed to solve hyperbolic 
systems of conservation laws (the so-called Godunov-type methods),
to calculate numerical solutions of the radiation transport 
equations in a static background. The feasibility of the method in any 
kind of regime, from diffusion to free-streaming, is demonstrated by a 
number of numerical tests and the effect of the choice of the closure
relation on the results is discussed.

\end{abstract}


%%%%%%%%%%%%%%%%%%%%%%%%%%%%%%%%%%%%%%%%%%%%%%%%%%%%%%%%%%%%%%%%
\section{Introduction}

In standard problems of radiation\footnote{ We will use the term 
{\it radiation} for both photons and neutrinos} hydrodynamics (RH)
where radiation contains a large fraction of the energy and momentum density,
the Boltzmann Equation (BE) must be coupled to the hydrodynamic 
equations in order to obtain the evolution of the system as well as the 
spectrum and angular distribution of the radiation field. 
However, an algorithm built to solve the BE
numerically (a Boltzmann solver) in a non-stationary case is too 
time-consuming, from a computational point of view, to allow for the 
extension to more than one dimension of the existing numerical codes 
\cite{MB93,Ya99}. In many cases, instead of the BE, its {\it angular 
moments} are considered, obtaining then the multigroup (or 
multi-frequency) equations or the even simpler energy averaged equations.

Standard RH methods \cite{MM84} have found their 
way into the literature, and one can find several RH codes with
different approaches for the radiation transport part, 
from single-energy-group 
(or two temperature) such as VISPHOT \cite{EB92} or TITAN \cite{GM94}, 
to multigroup radiative transfer, such as STELLA \cite{BB93}.    
Other authors built codes devoted more specifically to the radiation transport,
but at the expense of detailed hydrodynamics. An example is the 
code EDDINGTON \cite{EP93} in which free expansion is assumed.  
The hydrodynamics part of most of the existing codes is a one dimensional  
implicit finite difference scheme, including artificial viscosity terms   
for problems requiring an accurate treatment of shock waves and discontinuities.
During the last decade, a new subclass of numerical methods, the so-called 
Godunov-type methods, has been gradually substituting the schemes based 
on numerical viscosity due to their easier extension to multidimensional
cases and their greater capabilities in the treatment of strong shocks.  
The lack of a radiation transport method fully compatible with hydrodynamical 
Godunov-type schemes is one of the motivations of these paper. 

Turning back to the radiation transport part, another important  
issue related to the angular moment equations is the closure relation. 
By taking angular moments of the distribution 
function the complexity of BE is highly reduced, but the information about
the angular dependence of the radiation field is partially lost. Each 
$n^{th}$ angular moment equation of the BE contains angular moments of 
higher order, thus any truncated hierarchy of moment equations
contains more unknowns than equations and must be supplied with
additional equations or {\it closure relations} \cite{CB94,Gro96}.
Theoretically, if the 
correct closure for a given problem is known, the solution obtained for 
the first moments of the distribution function by solving the moment 
equations should be the same as the solution obtained by solving the BE.

Although the moment equations are much simpler to solve than the BE, 
in many situations an additional simplification is made by neglecting
some other terms, 
leading to the {\it diffusion approximation} (DA). In this approximation,
however, the resulting energy flux can be higher than the limit predicted 
by causality, specially in the regions where the radiation mean free path 
becomes comparable to the characteristic length. Different extensions of DA, 
like {\it flux-limited diffusion} 
\cite{BW82} or {\it artificial opacities} \cite{DJ92} have being used to 
overcome this problem. These extensions 
only partially solve the breakdown of the DA in the 
semi-transparent and transparent layers, and all of them are based on the 
same idea: to reintroduce some terms which had been dropped in the original 
assumptions of the DA. The term which is usually kept out of the equations 
is the time derivative of the fluxes. By neglecting this term, the character of 
the system of equations is changed to parabolic, and causality is therefore 
violated (disturbances  propagate at infinite velocity). In order
to develop a method which is valid in all regimes 
(from diffusion to free streaming) while preserving causality, one must solve 
the full set of equations keeping its hyperbolic character.

In this paper, we address the problem of establishing a well-defined 
hyperbolic system of equations for the first angular moments of the BE in 
the non-stationary spherically symmetric case. We will see that, besides 
the important question of its hyperbolicity, some constraints on the 
mathematical properties of the closure relations can be derived. They might 
help to disregard among some of the closures previously proposed. Moreover,
the behaviour of the characteristic fields of the hyperbolic system, 
which gives information on the velocity of perturbations, can also be used to
establish additional constraints to the closures. 

A direct consequence of considering the moment equations in 
hyperbolic form is that it permits the application of
powerful numerical techniques that have been developed
in recent years for hyperbolic systems of conservation laws (the equations of
classical and relativistic fluid dynamics for perfect fluids, for example).
Among the different numerical techniques, the so-called 
high resolution shock capturing (HRSC) methods 
have a number of interesting features such that stability,
being conservative, convergence to physical solutions and  
high accuracy in regions where the solution is smooth.
HRSC methods are based on the resolution of local Riemann problems 
(an initial value problem with discontinuous data) at 
the interfaces of numerical shells,
ensuring a consistent treatment of discontinuities and steep
gradients \cite{Go59}.
Their special relativistic extension \cite{MIM91} 
has shown its potential in simulations of heavy ion
collisions and extra-galactic jets, and different attempts
to extend the method to General Relativity have been done
\cite{BFIMM97,PFIMM98}.
We refer the interested reader to the recent reviews by
Mart\'{\i} \& M\"uller \shortcite{MM99} 
and Ib\'a\~nez \& Mart\'{\i} \shortcite{IM99}
for a detailed description of the current
status of HRSC techniques in numerical relativistic hydrodynamics.

Although HRSC are specially designed for hyperbolic systems without 
source terms which corresponds to the transparent regime, we will show in 
numerical experiments that, by appropriate
treatment of the sources, accurate solutions can be obtained also in the 
diffusion regime, where the source terms are dominant.

The structure of the paper is the following:
In \S 2 we summarise the deduction of the angular moment equations
of the BE in a spherically symmetric case, and for a static
background. We also describe the general form of the collision
terms when emission-absorption and iso-energetic scattering processes 
are included. In \S 3 we discuss the most common techniques used
to solve the transport equations, such as diffusion or flux-limited
diffusion, remarking their main features and limitations.
In \S 4 the hyperbolic character of the equations and its implications
for the closure relations are discussed. 
In \S5, the numerical techniques employed to solve the transport
equations as a hyperbolic system of conservation laws are discussed.
A number of numerical experiments solving the transport equations
in several test problems are displayed in \S 6,
and the feasibility of the
hyperbolic treatment in all kinds of regimes is demonstrated.
Finally, main conclusions
and advantages of our proposal are summarised in \S 7.

%%%%%%%%%%%%%%%%%%%%%%%%%%%%%%%%%%%%%%%%%%%%%%%%%%%%

\section{Transport equations in spherical coordinates}

We shall start our discussion from the radiative transfer 
equations in a static medium, deferring to a future work the 
inclusion of fluid velocity
terms in the equations. Although the restriction to zero velocity
seems to be specific, there are some astrophysical scenarios where  
the assumption that matter is at rest is a reasonable approach. 
Moreover, the inclusion of velocity terms, in some cases of interest
does not change the essentials of our conclusions.

In a static background, the Boltzmann Equation for massless particles in 
spherical coordinates can be written as follows\footnote{We work in units 
where $c=\hbar=1$.}

\begin{equation}
\label{BE}
\omega \left[ \frac{\partial {\cal I}}{\partial t} +
\mu \frac{\partial {\cal I}}{\partial r} +
\frac{(1- {\mu}^2)}{r} \frac{\partial {\cal I}}{\partial \mu} \right]
= \left( \frac {d{\cal I}}{d \tau} \right)_{coll}
\end{equation}
\noindent where ${\cal I}={\cal I}(t,r,\omega,\mu)$ is the invariant 
distribution function,
$\mu$ is the cosine of the angle of the particle momentum with respect
to the radial direction, $\omega$ the energy of the particle and the right 
hand side term is the invariant source or collision term coming from the interaction
between the radiation field and matter. 

As stated in the introduction, a common method of solving the BE is the 
method of moments
\cite{Tho81}, which involves 
taking angular moments of the equation by applying the operator

\begin{equation}
\frac{1}{4 \pi} \int_{-1}^{+1} \int_{0}^{2 \pi} 
{\mu}^{i}~d{\mu}~d\Phi \,\,\,\,\,\,\, i=0,1,2,... 
\end{equation}

Denoting by $E$, $F$ and $P$ the angular moments of 
the specific intensity $g \omega^3 {\cal I}$, $g$ being the 
statistical weight ($g=2$ for photons and $g=1$ for neutrinos)
\begin{eqnarray}
\label{defp}
E=E(t,r,\omega)=
g \frac{\omega^3}{2}\int_{-1}^{+1} d\mu ~{\cal I},
\nonumber \\
F=F(t,r,\omega)=
g \frac{\omega^3}{2}\int_{-1}^{+1} d\mu~\mu ~{\cal I},
\nonumber \\
P=P(t,r,\omega)=
g \frac{\omega^3}{2}\int_{-1}^{+1} d\mu~\mu^2 ~{\cal I},
\end{eqnarray}
the equations 
corresponding to the first two moments (0 and 1) of the BE are:

\begin{eqnarray}
\label{mom0}
\partial_t E  + \partial_r F + \frac{2 F}{r} = s^0
\end{eqnarray}
\begin{eqnarray}
\label{mom1}
\partial_t F  +  \partial_r P + \frac{ 3P-E}{r} = s^1
\end{eqnarray}
where the moments of the collision term are defined as
\begin{equation}
\label{q0}
s^0 = g \frac{{\omega}^2 }{2} \int_{-1}^{+1}
{\left(\frac {d{\cal I}}{d \tau}\right)}_{coll} d{\mu}
\end{equation}
\begin{equation}
\label{q1} 
s^1 = g \frac{{\omega}^2 }{2} \int_{-1}^{+1}
{\left(\frac {d{\cal I}}{d \tau}\right)}_{coll} \mu d{\mu}
\end{equation}

Notice that, in the absence of velocity fields or strong gravitational 
fields, the different energy groups in a multifrequancy scheme are    
only coupled through the source terms, rendering the mathematical     
character of the left hand side of equations (\ref{mom0})-(\ref{mom1})
formally identical to the energy-averaged problem. 

%%%%%%%%%%%%%%%%%%%%%%%%%%%%%%%%%%%%%%%%%%%%%%%%%%%%

\subsection{Collision terms}

The most general form for the collision term including emission, absorption
and isoenergetic scattering is the following:

\begin{equation}
{\left(\frac {d{\cal I}}{d \tau}\right)}_{coll} = 
\omega\left(j - \kappa_a {\cal I} + \kappa_s[{\cal I}]\right)
\end{equation}
where $j$ is the emissivity, $\kappa_a$ the absorption
opacity including final states blocking (for fermions) or
stimulated emission (for bosons) and $\kappa_s$ the scattering opacity  related
to the scattering
reaction rate ($R^s$) through
\begin{equation}
\kappa_s(\omega) = { \left( \frac{\omega}{2 \pi} \right)}^3 \int_{-1}^{+1} 
d \mu' \int_0^{2 \pi} d \varphi \, R^s(\omega,\cos \Theta) \, 
({\cal I}(\omega, \mu')- {\cal I}(\omega, \mu))
\end{equation}
where $\Theta$ is the angle between the in-going and outgoing particle
and $\varphi$ is the azimuthal angle. The opacities and the emissivity
are expressed in units of inverse length.

For isotropic scattering the source terms appearing in equations 
(\ref{mom0}) and (\ref{mom1}) can be written as
\begin{equation}
s^0 = \kappa_a  \left( E^{eq}-E \right)
\label{S0}
\end{equation}
\begin{equation}
s^1 = - \kappa  F
\label{S1}
\end{equation}
where $ \kappa = \kappa_a + \kappa_s $
and  $E^{eq}$ is the value of $E$ in equilibrium with matter.
In the above, we have assumed matter in local thermodynamic equilibrium (LTE),
and only the radiation field (photons or neutrinos) is allowed to deviate 
from equilibrium. 
The conclusions of our study might be affected                      
by large velocity fields or for frequencies close to line               
discontinuities, which are present in some astrophysical scenarios, and
deserve a more careful study \cite{MSKH76,Ku83,MK86}.

%%%%%%%%%%%%%%%%%%%%%%%%%%%%%%%%%%%%%%%%%%%%%%%%%%%%

\section{Flux-limited diffusion and artificial opacities}
One of the approaches most widely used to solve the transport
equations numerically is the {\it diffusion approximation}, in which 
the invariant distribution function is assumed to be nearly isotropic 
in the comoving frame.
Thus an expansion in terms of Legendre polynomials
to the order $O(\mu^2)$ is enough to maintain the main features of the
radiation field
\begin{equation}
{\cal I}(\mu)= {\cal I}_0 + 3~{\cal I}_1~\mu
\end{equation}
where ${\cal I}_1 \ll {\cal I}_0$. Consistent with this 
assumption, the time derivative of the flux is neglected in equation 
(\ref{mom1}), and together with equation (\ref{q1}) 
gives the following relation
for the flux in terms of the energy gradient
\begin{equation}
\label{f1}
F(\omega) = - \frac{1}{3 \kappa(\omega)}
\frac{\partial E(\omega)}{\partial r}
\end{equation}

As stated before, the DA breaks down when the mean free path is large 
compared to the typical scale of the problem, and the fluxes calculated 
with the former formula may give non-causal values, in the sense that
the flux can be greater than the energy density. 
This pathology comes from the fact that we have
neglected some terms in the momentum equation in order to obtain a simple 
formula for the fluxes. 

Flux limiters have been introduced {\it ad hoc} in 
the literature to avoid this non-causal behaviour; see, {\it e.g.}, Minerbo 
\shortcite{Mi78}, Levermore \& Pomraining \shortcite{LP81} or Cernohorsky 
\& Bludman \shortcite{CB94}. In order to illustrate where the problems stem 
from, we begin by deriving the {\it flux-limited  diffusion}
equations. 
Let us define the {\it flux factor}, $f$, and the {\it Eddington factor}, $p$:
\begin{equation}
f = \frac{F}{E}
\end{equation}
\begin{equation}
p = \frac{P}{E}
\end{equation}
By subtracting equation (\ref{mom0}) multiplied by $f$ from (\ref{mom1}), and
after some algebra (now keeping all terms), one obtains:
\begin{equation}
F = - \frac{(p - f^2)}
{(\kappa_{ph}+\kappa_{J})} \partial_r E
\label{Fdif}
\end{equation}
where the quantities, including the flux and Eddington factors, are now energy 
dependent. The explicit expression for the different {\it opacities} are
\begin{equation}
\kappa_{ph} = \kappa_s
+ \kappa_a \frac{E^{eq}}{E}
\end{equation}
\begin{eqnarray}
f \kappa_{J} = {\partial_t f} + {\partial_r p} - f \partial_r f
+ \frac{(3p-1-2f^2)}{r}
\end{eqnarray}

The {\it physical opacity}, $\kappa_{ph}$, or effective albedo only depends 
on the interaction of radiation with matter and includes an 
out--of--equilibrium correction to absorption opacity. The second 
term in the denominator, $\kappa_J$, is the multigroup 
extension of the so-called {\it artificial opacity} \cite{DJ92} and takes into 
account geometrical corrections and deviations from the diffusion limit
(it vanishes in the limit $f \ll 1$).

The term in the numerator of equation (\ref{Fdif}), $(p - f^2)$, plays the role 
of a flux--limiter. It has the value $\frac{1}{3}$ in the diffusion
limit and goes to zero as the free streaming limit ($f \rightarrow 1$
, $p \rightarrow 1$) is reached.
To close the set of equations, a closure relation for $p$
is needed, and it is directly related to flux limiter $(p-f^2)$.
%Giving a closure is the price
%one must pay to avoid solving the BE, which is more
%difficult technically and much more expensive computationally.
We will re-address the closure relation problem in next section.

With this description of the transport equations, the simple structure 
of the diffusion equation and its parabolic character is maintained when the 
first term in $\kappa_J$ is neglected ($\partial_t f$ = 0).
The geometric and velocity field corrections can also be included
through artificial opacities as corrections to the physical opacity.
Although the flux-limiter helps to avoid non-causal behaviour in the 
sense $F>E$, causality is also violated when the time derivative of 
$f$ is neglected, because the character of the equations change to 
parabolic, which results in an infinite speed of propagation 
of the information. By keeping this term, the diffusion scheme is, in some 
applications, numerically unstable and further numerical tools are needed 
to handle these instabilities. This is the subject of the next section.

%%%%%%%%%%%%%%%%%%%%%%%%%%%%%%%%%%%%%%%%%%%%%%%%%%%%

\section{Hyperbolic Formulation of Transport Equations}

The system of equations (\ref{mom0}) and (\ref{mom1}) can be
written in a compact form as follows
\begin{equation}
\label{system}
\frac{\partial \vec{\cal U}}{\partial t}
+ \frac{1}{r^2} \frac{\partial \left( r^2 \vec{\cal F} \right)}
{\partial r} = \vec{\cal S}
\end{equation}
where the vectors $ \vec{\cal U}$ and $\vec{\cal F}$ are
\begin{equation}
\vec{\cal U} =(E,F)
\end{equation}
\begin{equation}
\vec{\cal F} = (F, pE)
\end{equation}
and the sources $\vec{\cal S}$, according to equations (\ref{S0}-\ref{S1}),
are given by
\begin{equation}
\vec{\cal S} = \left( \kappa_a (E^{eq}-E), -\kappa F + \frac{E-P}{r} \right).
\end{equation}
Notice that the sources are free of partial derivative operators.

The above system of equations (\ref{system}) is said to be
a hyperbolic system of 
conservation laws (with source terms) when the Jacobian matrix 
associated to the {\it fluxes} ,$\nabla_{\vec{\cal U}} \vec{\cal F}$, has 
real eigenvalues 
and there exists a complete set of corresponding eigenvectors.
To analyse the Jacobian matrix, we will assume that the Eddington factor
is only a function of the flux factor $p=p(f)$. This assumption is commonly
used by a number of authors who have proposed different closure relations or
flux-limiters in the context of flux-limited diffusion.
In this case, the Jacobian matrix is
\be
\nabla_{\vec{\cal U}}\vec{\cal F}
= \left ( \matrix{
& 0              & 1      \cr
& p - f p' & p'  \cr } \right )
\ee
where  $p' = \displaystyle{\frac{d p}{d f}}$ .

The eigenvalues of the above matrix are
\begin{equation}
\label{lambda}
\lambda_{\pm} = \frac{p' \pm \sqrt{ {p'}^2 + 4(p - f p') } }{2}
\end{equation}
and the right eigenvectors are
\begin{equation}
\label{vector}
\vec{r}_{+} =  (1,\lambda_{+}) ; ~~~  \vec{r}_{-} =  (1,\lambda_{-})
\end{equation}
The mathematical character of the system of equations 
(\ref{system}) depends on the value of the discriminant 
in equation (\ref{lambda}), which can be written as follows
\be
\label{disc}
\Delta = {({p'}-2f)}^2 + 4(p -f^2). 
\ee
Since $p$ and $f$ are normalised moments of a non--negative weight 
function, on the unit sphere, there is an extra relation between $f$ 
and $p$, the {\it Schwarz inequality}
\be
\label{schw}
f^2 \le p \le 1 ~~~, 
\ee
that ensures that for
any closure satisfying (\ref{schw}) the eigenvalues 
are always real and hence the hyperbolic character is guaranteed. In addition, 
causality imposes that the velocity of propagation of perturbations must be 
smaller than the speed of light, which gives another constraint on the 
eigenvalues
\be
\left| \lambda_{\pm} \right| \le 1 ~,
\ee
that can be expressed in terms of $f$, $p$ and $p'$ as follows
\be
-\frac{1-p}{1+f} \le p' \le \frac{1-p}{1-f}.
\label{ppb}
\ee
This restriction, together with equation (\ref{schw}) and the
condition $|f| \le 1$,
defines the region in the $[f,p,p']$ space allowed for causal closures.

In many astrophysical problems, radiation diffuses out from a 
central opaque region to an outer optically thiner region. In such a 
scenario, the radiation angular distribution is nearly isotropic in
the central region  ($|f| \ll 1$) and forward peaked in the transparent 
region, far from the opaque centre ($f \rightarrow 1$). Thus, the closure
in that problem must satisfy 
\be
p(f=0)=\frac{1}{3} \quad {\rm and} \quad p(f=1) = 1.
\ee
The first limit comes from the diffusion approximation while the
second limit comes from the Schwarz inequality (\ref{schw}), and both
are satisfied by all the closures found in the literature (see Table I).
Another general property is that, in the diffusion regime
the Eddington factor has the constant value
of $1/3$, thus another restriction on the derivative can be imposed
\be
\label{cons2}
p'(f=0) = 0 ~,
\ee
that leads to the characteristic speeds 
$\lambda_{\pm} = \pm \frac{1}{\sqrt{3}}$.
In the limit $f=1$, equations (\ref{schw}) and (\ref{ppb}) gives the bounds
$ 0 \le p' \le 2 $ and
the eigenvalues in this limit are $\lambda_{+}=1$ and $\lambda_{-}=p'-1$.
 According to the theory of hyperbolic systems, this result can be 
interpreted as a first wave propagating outwards
at the speed of light and a second wave with a characteristic speed that
depends on the value of $p'$. 
Although  $\lambda_{-}$ depends on the form of the closure, for  $f=1$
the amplitude of this second wave is zero and the value of $p'$ does not
affect the evolution. However, for $f$ close but not equal to one,
the choice of $p'$ can be important, as we will show in section \S 5.2.
Similar conclusions can be obtained for the case $f=-1$.

In table 1 we summarise the main properties of some of the most widely 
used closures: Minerbo \shortcite{Mi78}, LP \cite{LP81},
the linear formula by Auer \shortcite{Auer84}, 
Kershaw's parabolic approach \cite{Ke76}, the flux-limiters used by Bruenn
\shortcite{Bru85} and Bowers \& Wilson \shortcite{BW82}, Janka's fit 
to Montecarlo simulations of neutrino transport in Supernovae \cite{Jan91} 
and that corresponding to an isotropically emitting sphere \cite{CVB86}, 
named {\it vacuum}.

There are several comments about Table I that we would like to address.
First, one closure relation (BW) leads to
non-causal values of one eigenvalue in the diffusion regime.
Secondly, four closures do not satisfy the condition  $p'(f=0)=0$
(Auer, Bruenn, BW, Vacuum). 
Auer's formula failure is due to an excessive simplification (it is
a linear approach) while 'Vacuum' has been obtained from the assumption
of an isotropically radiating sphere, which restricts their validity
to $f \ge 0.5$.
The closure relations proposed by Kershaw, Janka, Minerbo and LP have the
common features of preserving causality leading to  
reasonable values of the characteristic velocities 
($\lambda_{\pm} = \pm \frac{1}{\sqrt{3}}$) in the diffusion regime.
Thirdly, it is worthy to point out that all the closures
except Auer's give $p'\ge 1$
when approaching the radially-streaming limit $(f=1)$, which makes
both eigenvalues be positive. It ensures that no information can be
propagated inwards for sufficiently high values of $f$.
It is clear that these closures cannot be used to treat problems
in which a small perturbation is propagating inwards from the transparent
region, as we will show in next section.

%-------------------------------------------------------------------

In a more general situation, the Eddington factor depends on both, 
the flux factor and the occupation density corresponding to the zeroth 
angular moment of the distribution function, 
$e=E/\omega^3 $. 
The Jacobian matrix associated to the fluxes is

$$\frac{\partial \vec{\cal F}}{\partial \vec{U}}
=  \left ( \matrix{
& 0       &      & 1      \cr
&         &      &        \cr
& p - f {\displaystyle \left. \frac{\partial p}{\partial f} \right|_e}+ 
e {\displaystyle \left. \frac{\partial p}{\partial e} \right|_f}
& & {\displaystyle \left. \frac{\partial p}{\partial f} \right|_e } \cr 
} \right)  $$
and now the eigenvalues are
\begin{equation}
\lambda_{\pm} = \frac{1}{2} \left(
\left. {\displaystyle \frac{\partial p}{\partial f}} \right|_e
\pm \sqrt{\Delta} \right)
\end{equation}
where the discriminant is
\begin{equation}
\Delta = {\left(  \left. \frac{\partial p}{\partial f} \right|_e 
- 2~f \right)}^2 + 4 \left( p - f^2 + 
e \left. \frac{\partial p}{\partial e} \right|_f \right)
\end{equation}

In contrast to (\ref{disc}), the sign of the discriminant might be
negative for negative values of  
$\left. \frac{\partial p}{\partial e} \right|_f$, and the system
would become elliptic instead of hyperbolic. 
As stated before, hyperbolic 
equations are related to finite speeds of propagation of the 
perturbations, while elliptic equations are related with boundary
problems and do not allow for solutions
with propagating signals. This fact indicates that, in order to
preserve hyperbolicity, only closures which 
give positive values of the discriminant should be considered. 

A closure of this kind has been proposed by 
Cernohorsky \& Bludman \shortcite{CB94} using arguments of maximum
entropy. Minerbo's, LP's and the Vacuum closures can be obtained as 
limiting cases of CB closure
in the low occupation limit (classical statistics), large ocuppation limit
for Bose-Einstein statistics and maximal forward packing limit for 
Fermi-Dirac statistics, respectively. We have checked that CB closure
gives real and causal eigenvalues for both, Bose-Einstein and Fermi-Dirac
statistics, in any regime.

%%%%%%%%%%%%%%%%%%%%%%%%%%%%%%%%%%%%%%%%%%%%%%%%%%%%%%%%%%%%%%%
\section{The numerical approach.}

The numerical code employed for the calculations presented in 
\S 6 is a finite finite difference scheme in which 
the left (L) and right (R) states for the Riemann problems set
up at each interface have been obtained
with a monotonic, piecewise linear reconstruction procedure 
\cite{vL79}. The numerical fluxes are evaluated following the idea
of Harten, Lax \& van Leer \shortcite{HLL83}, that is based on approximate
solution of the original Riemann problem with a single intermediate state.
The resulting numerical flux is given by
\be
\hat{\cal F}^{HLL} = \frac
{\psi^+ {\cal F}_L - \psi^- {\cal F}_R + 
\psi^+  \psi^- ({\cal U}_R - {\cal U}_L)}{\psi^+ - \psi^-}
\ee
where $\psi^+ = \mbox{max}(0,\lambda_+^R, \lambda_+^L)$ and
$\psi^- = \mbox{min}(0,\lambda_-^R, \lambda_-^L)$.

The time-advance algorithm is a TVD-preserving, third order Runge-Kutta
that takes into account the influence of the source terms in intermediate 
steps.  The source term in this problem deserves a special mention.
It acts as a relaxation term which
leads to a long-time behaviour governed by the reduced parabolic
system described in section \S 3. Although high resolution methods
for hyperbolic conservation laws usually fail to capture this feature 
(unless a very fine grid is used), 
the use of a modified flux \cite{JL96}
and an implicit linearization of the source term in each Runge-Kutta
step allows us to succeed in the treatment of such stiff relaxation term 
with coarse grids and
time-steps longer than the source time-scale ($\approx 1/\kappa$),
as we will show in subsection \S 6.1.
Although we have constrained the numerical experiments to situations
when only scattering processes are permitted, the inclusion of
absorption-emission processes can be treated in a similar way. However
it introduces energy and momentum exchange with the matter and makes
necessary to solve the equations of radiation transport coupled with
the equations of hydrodynamics or the equations of hydrostatic equilibrium
with number and heat transfer.
Such a problem is out of the scope of this paper, in which we focussed
on the radiation part, and will be addressed in forthcoming works.

%%%%%%%%%%%%%%%%%%%%%%%%%%%%%%%%%%%%%%%%%%%%%%%%%%%%%%%%%%%%%%%
\section{Numerical experiments and discussion.}

To check the applicability of HRSC techniques to radiation
transport problems, we have performed a number of numerical tests  
considering several simplified models in which the 
analytical solution is known. In the remainder of this section 
we describe the different tests and display the results.
Our intention is to prove that our method can solve
radiation transport in all kinds of regimes, from diffusion (TEST 1)
to radially streaming (TEST 2), going through the semi-transparent
region (TEST 5) and in situations where strong gradients are
developed (TEST 4).

Ideally, in a multi-group scheme we have as many systems of two equations 
as energy bins
in which the radiation density is splitted. In a radiation hydrodynamics
problem all those pairs of equations are coupled to the matter equations
and must be solved together. Since our objective is to study the 
applicability of HRSC techniques for the radiation part, in what follows
we assume to have only one energy group or, equivalently, we consider 
the energy integrated equations. 
Notice, though, that the conclusions
are not affected since, when velocity fields are not taken
into account, the equations for different energies are
only coupled through the source terms. 
A different problem will arise when velocity fields are important or 
gravitational and Doppler red-shift are considered. In those situations
either the anisotropy of the source term (in the observer frame) or
the advection and aberration terms (in the comoving frame) cause
the energy redistribution along the photon (or neutrino)
path and a more careful treatment needs to be done.

%%%%%%%%%%%%%%%%%%%%%%%%%%%%%%%%%%%%%%%%%%%%%%%%%%%%%%%%%%%%%%%%%%%%%%%%
\subsection{TEST 1: Diffusion limit}

In the diffusion limit, the closure is obviously $p=1/3$
and the time derivative of the flux can be neglected, leading to 
equation (\ref{f1}). 
Setting the absorption opacity to zero and
the scattering opacity to a constant value, the analytical solution
of the system formed by (\ref{mom0}) and (\ref{f1}), corresponding to 
an initial value problem consisting
in a Dirac delta located in $r=0$ at $t=0$, is the following
\be
E(r,t) = {\left( \frac{\kappa}{t} \right)}^{3/2} 
\exp \left\{ - \frac{3 \kappa r^2}{4t} \right\}
\ee
\be
F(r,t) = \frac{r}{2t} E(r,t)
\ee
We imposed $F(r=0)=0$ as inner boundary conditions and outflow 
boundary conditions in the outermost shell.

\begin{figure}
\psfig{figure=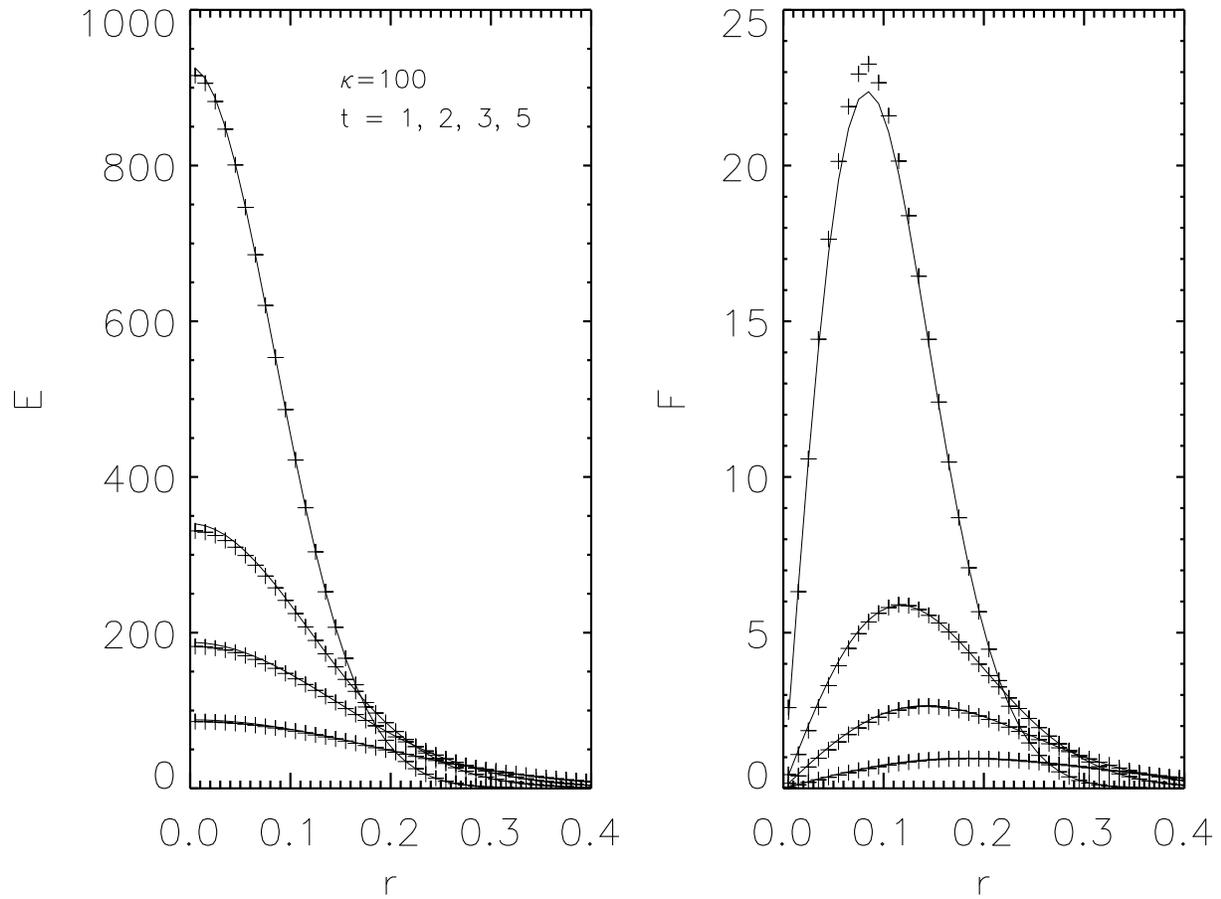,width=16cm}
\caption{Moderate diffusion limit: Energy density (left panel) and 
flux (right panel) as a function
of the radial coordinate for different times, ($t=1,2,3,5$ from top
to bottom). The opacity has been set to $\kappa=100$, corresponding
to a Peclet number of $Pe=1$.}
\label{fig1a}
\end{figure}

\begin{figure}
\psfig{figure=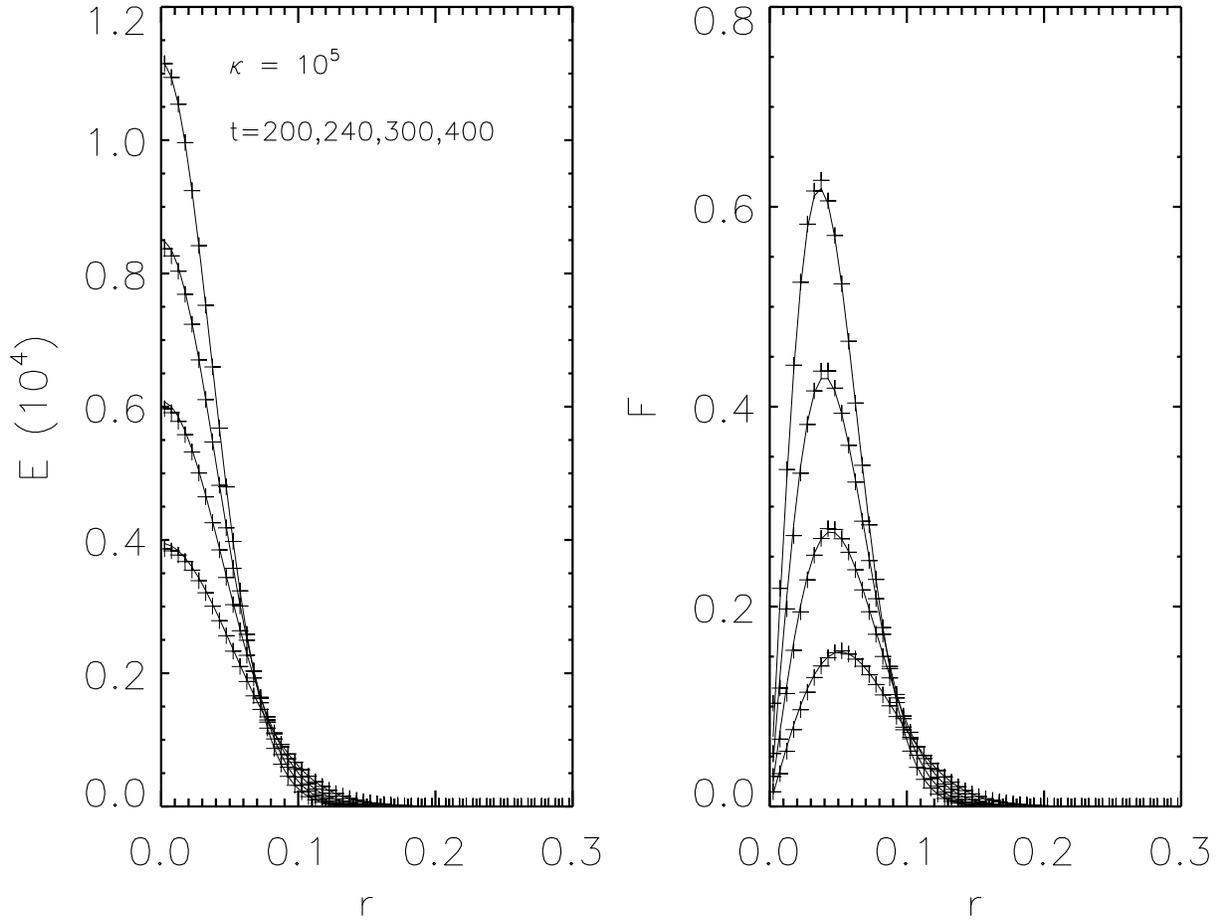,width=16cm}
\caption{Extreme diffusion limit: Energy density (left panel) and 
flux (right panel) as a function
of the radial coordinate for different times, ($t=200,240,300,400$ from top
to bottom). The opacity has been set to $\kappa=10^5$, corresponding
to a Peclet number of $Pe=5 \times 10^4$.}
\label{fig1b}
\end{figure}

We have performed two simulations. First, we set a value of $\kappa=100$
on a equally spaced grid of 100 shells in the domain $[0,1]$. It
corresponds to a Peclet number ($Pe=\kappa \Delta x$) of order unity.
The cell {\it Peclet number} is a measure of the spatial resolution relative
to the relaxation scale and a grid is coarse when $Pe \gg 1$.
This first simulation starts at $t=1$ to avoid
the numerical problems produced by an infinite value in $t=0$.
Results are shown in Figure \ref{fig1a}. 
The energy density (left panel) and the flux (right panel) are displayed for 
different times in dimension-less units, $t=1,2,3,5$ from top to bottom.
The solid line corresponds to the analytical solution and the crosses
correspond to the numerical solution.
Notice that, although we are finding numerical solutions of 
system (\ref{system}), the analytical solution of the diffusion limit 
is well reproduced by the numerical code, since
the opacity is large enough to be close to the diffusion limit.
The second simulation has been performed in a grid of 100 shells in
the domain $[0,0.5]$, and using a value of $\kappa=10^5$, which
corresponds to $Pe=5 \times 10^4$. The initial time in this case is 200
and the Courant time-step is of the same order ($10^{-2}$) as in the previous
simulation, three orders of magnitude larger than the relaxation scale. 
Results are displayed in Fig. \ref{fig1b}. The agreement between the
analytical and the numerical solutions clearly proofs the capability
of the code to handle situations with very stiff source terms. 

%%%%%%%%%%%%%%%%%%%%%%%%%%%%%%%%%%%%%%%%%%%%%%%%%%%%%%%%%%%%%%%%%%%%%%%%
\subsection{TEST 2: Radially Streaming limit}

\begin{figure}
\psfig{figure=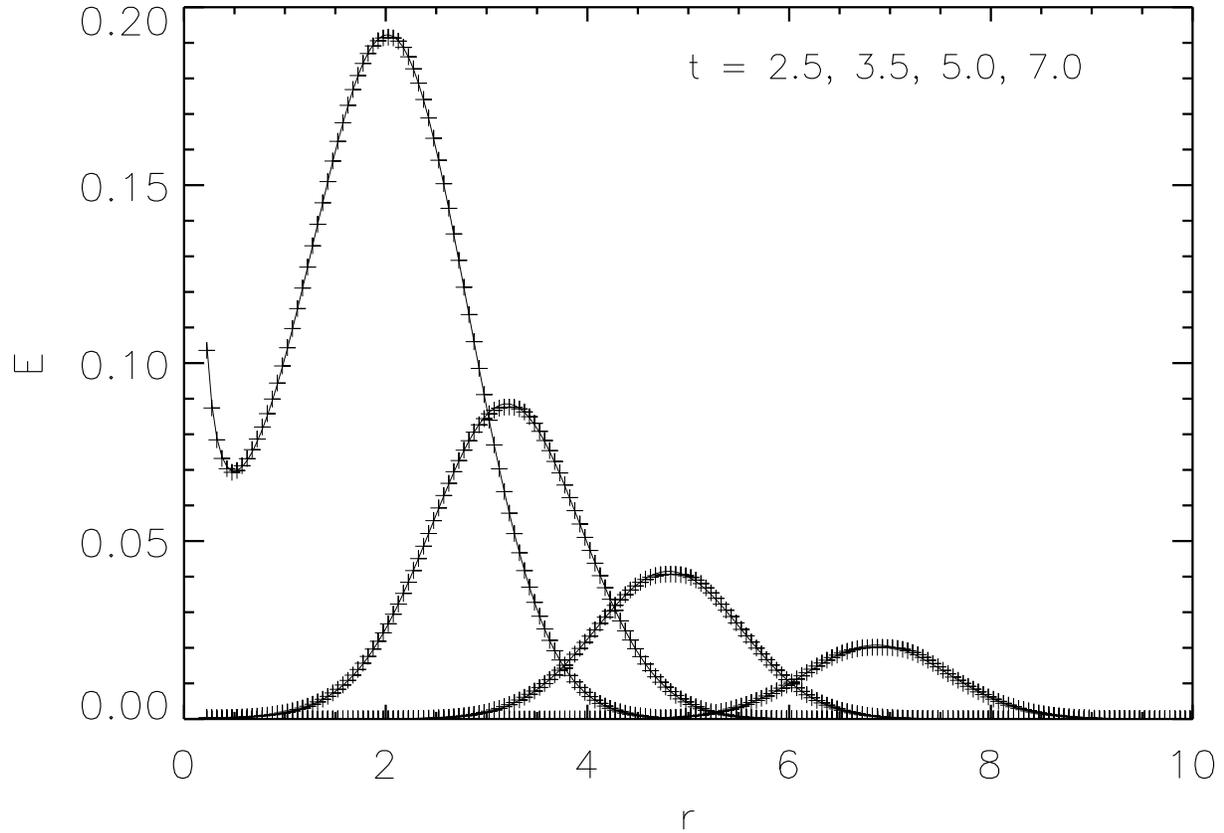,width=16cm}
\caption{Free streaming limit: A Gaussian spherical wave propagating outwards.
The analytical solution is represented by the solid line and the numerical
solution by crosses. Snapshots for four different times ($t=2.5,3.5,5.0,7.0$)
are plotted.}
\label{fig2a}
\end{figure}

In the radially streaming limit ($|f|=1$, $p=1$), and assuming there is
no interaction between radiation and matter ($\kappa_a=0$, $\kappa=0$)
equations (\ref{system}) are reduced to
\begin{eqnarray} 
\frac{\partial E(\omega)}{\partial t} +
\frac{1}{r^2} \frac{\partial (r^2 E(\omega))}{\partial r} 
= 0 
\end{eqnarray} 
The solution to this equation is any linear combination of
ingoing and outgoing spherical waves, {\it i.e.}, any linear
combination of functions of the form 
$E(t,r)=\frac{1}{r^2}
g(r \pm t)$. For the first numerical experiment we choose a Gaussian
spherical wave propagating outwards, 
\be
E(r,t) = \frac{1}{r^2} \exp \left\{ - {(r-t)}^2 \right\}
\ee
\be
F(r,t) = E(r,t).
\ee
Our numerical grid consists on 200 equally spaced shells 
going from $r_1=0.2$ to $r_2=10.2$ and the boundary conditions
consist in imposing the analytical solution in both boundaries.
We show the results of this test in Figure \ref{fig2a}. In the figure
we displayed the analytical solution (solid line) and the 
numerical (crosses) for the energy density. 
In order to analyse the effect of the particular choice for the closure, 
we repeated the experiment using different values for $p'(f=1)$, which leads
to different speeds of propagation of the waves, and no difference
was observed. The final reason is that all the closures displayed in Table 1
give $\lambda_+ (f=1)=1$. This is a general property of any
closure, since $p-f^2$ is a positive defined function
that goes to zero as $f \rightarrow 1$ and, therefore, its derivative has to
be negative in a vicinity of $f=1$ and $p' \le 2f$. Thus taking
the limit $f \rightarrow 1$ in (\ref{lambda}), one obtains
\be
\lambda_+ = \frac{p'+\sqrt{(2-p')^2}}{2} = 1
\ee

To understand why the value of $\lambda_-$ is not important in this
problem, we study the diagonalized system obtained by adding and
subtracting the two equations in (\ref{system}) after taking $p=1$,
\begin{eqnarray}
\frac{\partial (E+F)}{\partial t} +
\frac{1}{r^2} \frac{\partial (r^2 (E+F))}{\partial r} = 0
\\
\frac{\partial (E-F)}{\partial t} -
\frac{1}{r^2} \frac{\partial (r^2 (E-F))}{\partial r} = 0.
\end{eqnarray}
In this equations it is clearly seen that
the quantities propagated with velocities $\lambda_{\pm}$ (Riemann invariants)
are $E \pm F$, respectively. Since in our particular problem
we took $F=E$, the quantity propagated by $\lambda_-$ is 
identically null and the value of the speed of propagation
does not affect the solution.

\begin{figure}
\psfig{figure=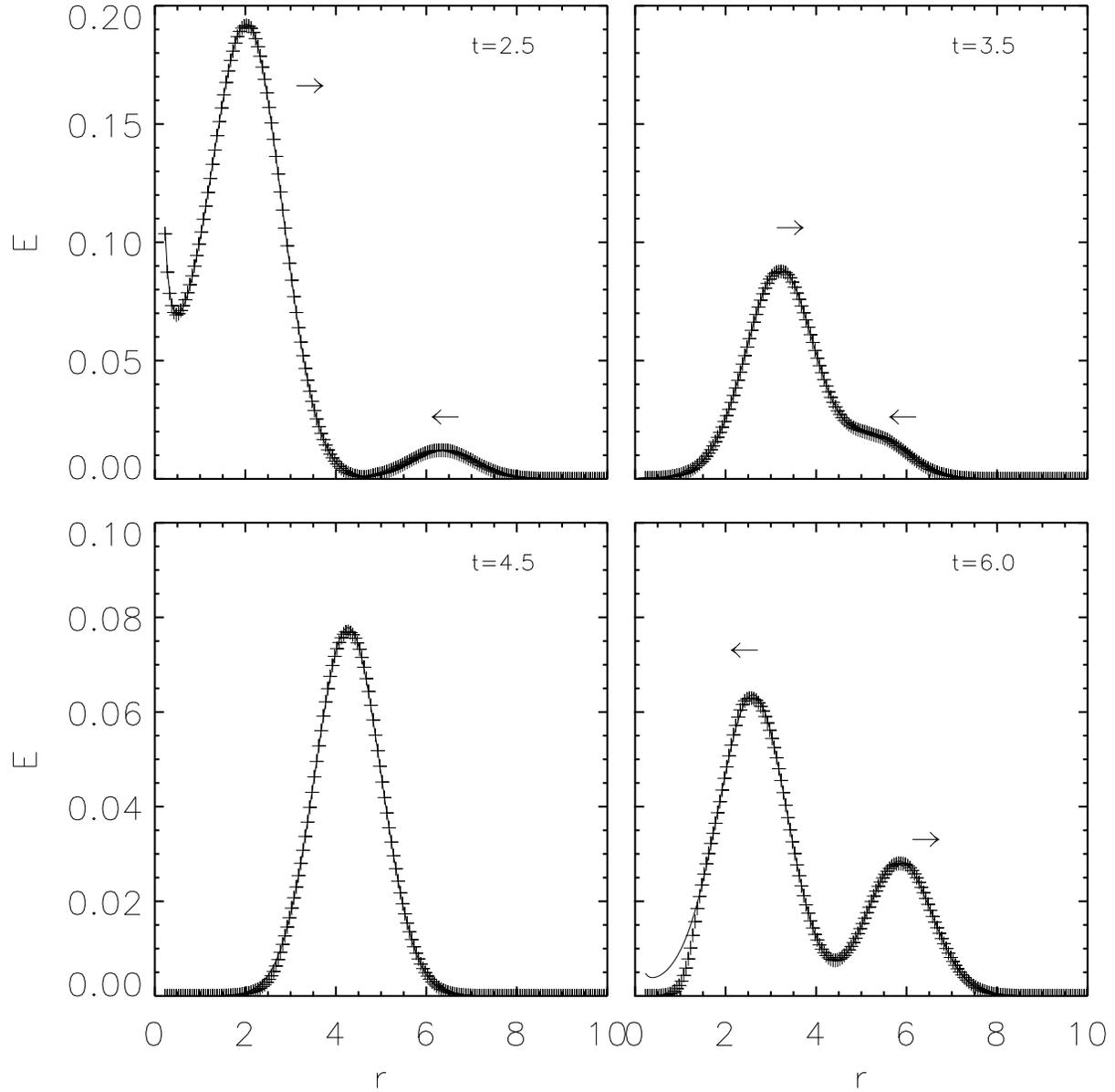,width=16cm}
\caption{Free streaming limit: Two Gaussian spherical waves crossing over.
The solid line corresponds to the analytical solution and the crosses
to the numerical solution. Each panel displays a snapshot for a 
different time ($t=2.5,3.5,4.5,6.0$). The closure employed is $p=1$,
$p'=0$, which gives $\lambda_{\pm}=\pm 1$
everywhere.}
\label{fig2b}
\end{figure}

A different situation is found when we add to the initial conditions
a second wave propagating in the opposite direction. Figure \ref{fig2b}
shows the results from such experiment, in which a second wave
with smaller amplitude and opposite direction is located initially at $r=7$.
The closure used was $p=1$, $p'=0$, which gives $\lambda_+=1$, $\lambda_-=-1$.
In the figure, the analytical solution is denoted by the solid lines and 
the numerical solution by crosses and each panel corresponds to a different
time. It can be seen how the code is able to solve the interaction
of the two waves until they completely cross over.
The difference between the analytical and numerical solution near
the center for the last panel are due to the boundary conditions
(initially, it was an ingoing wave).

\begin{figure}
\psfig{figure=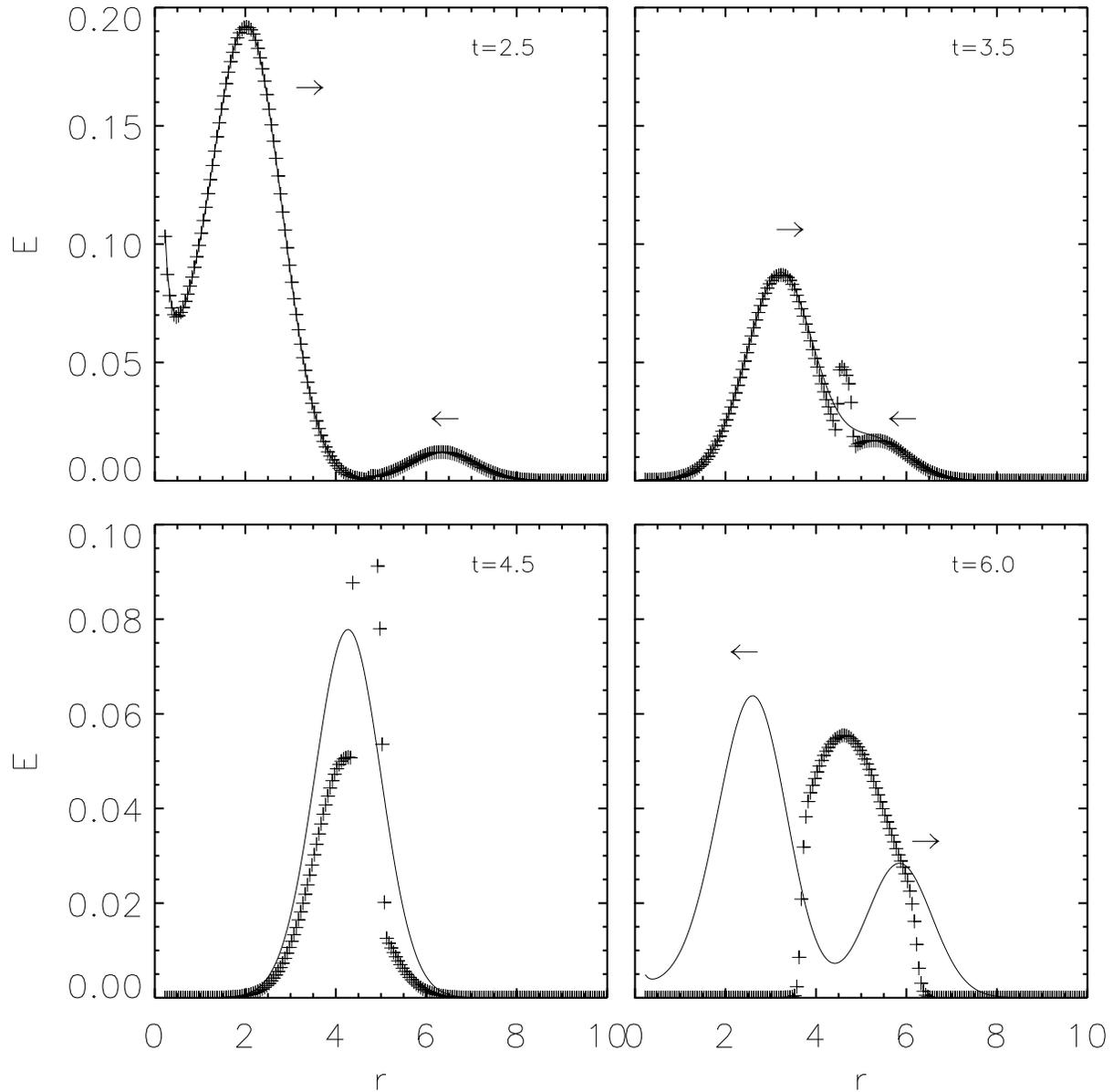,width=16cm}
\caption{Same as Figure \ref{fig2b} but using Kershaw's closure.
The effect of taking wrong characteristic speeds is clearly
illustrated.}
\label{fig2c}
\end{figure}

To illustrate the effect of a wrong speed of propagation, the
experiment is repeated with Kershaw's closure, that gives a correct speed
of propagation when $f=\pm 1$ but a wrong value otherwise.
Results are shown in Figure \ref{fig2c} for the same time steps
as in the previous case. At the beginning both waves seem to propagate
correctly but, as they cross each other and the flux factor departs
from unity, the numerical result differs from the analytical solution.
A similar behaviour is obtained with other closures.
This simple example manifests the importance of the choice for
the closure, since it fixes the speeds of propagation of the waves.

%%%%%%%%%%%%%%%%%%%%%%%%%%%%%%%%%%%%%%%%%%%%%%%%%%%%%%%%%%%%%%%%%%%%%%%%
\subsection{TEST 3: Sphere radiating in the vacuum.}

\begin{figure}
\psfig{figure=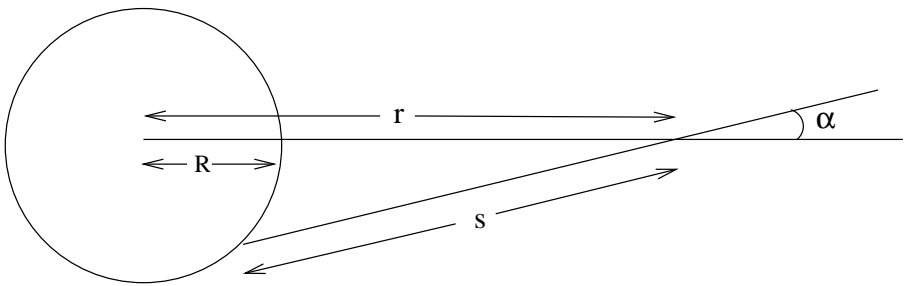,width=12cm}
\caption{Schematic diagram of Test 3: Sphere radiating isotropically
into vacuum. The analytical solution for the radiation distribution
function can be obtained in terms of $r$ and $\mu= \cos \alpha$.}
\label{vac}
\end{figure}

In the two previous tests we studied the limiting cases, diffusive
and free-streaming regimes.
Another academic problem in the limit of vanishing
opacities consists in a sphere of radius $R$ radiating isotropically into 
vacuum (Figure \ref{vac}). 
Given the explicit form of the distribution function at the
surface of the sphere, ${\cal I}_0(t)={\cal I}(t=0,r=R)$, the general solution 
can be easily obtained
 
\be
{\cal I}(t,r,\mu) = \left\{ {\matrix{
{\cal I}_0(t-s) &  \mu \ge x(r,t) \cr
0        &  \mu <  x(r,t) \cr}} \right.
\ee
where 
\be
s=r \left( \mu - \sqrt{\mu^2-1+{(R/r)}^2} \right)
\ee
\be 
x(r,t) = \left\{ {\matrix{
\sqrt{1-{(R/r)}^2} & t \ge \sqrt{r^2-R^2} \cr
\displaystyle{ \frac{t^2+r^2-R^2}{2tr}} & r-R<t<\sqrt{r^2-R^2} \cr
1     &  t < r-R \cr}} \right.
\ee
where $\mu=\cos \alpha$.

We take a linear time dependence ${\cal I}_0(t)=C t$, being $C$
an arbitrary constant, so that the angular integrals can be done 
analytically.  The angular moments of the distribution
function are, then, given by
\be
E(t,r)=\frac{C}{4} \left[ 2t(1-x) -r (1-x^2) + r \sqrt{1-x^2}
+ r x^2 \log \left( \frac{x}{1 + \sqrt{1-x^2}} \right)   \right]
\ee

\be
F(t,r)=\frac{C}{12} \left[ 3t(1-x^2)
-2r (1-x^3) + 2r (1-x^2)^{3/2} \right]
\ee

\be
P(t,r)=\frac{C}{48} \left[ 8t(1-x^3) - 6r (1-x^4) + 3r (2-x^2) \sqrt{1-x^2}
+ 3 r x^4 \log \left( \frac{x}{1 + \sqrt{ 1-x^2}} \right)   \right].
\ee

We set up a numerical grid of 200 points, uniformly distributed,
from $r=1$ to $r=200$. The initial conditions are $F=0$ and
$E=\alpha/r^2$ with $\alpha$ being a small number, different from 
zero for numerical purposes. We have taken $\alpha=10^{-4}$. 
The radiation field is
introduced through the inner boundary conditions,
$E(t,r=R)={2t}$ and $f(t,r=R)=0.5$. Outflow is assumed in 
the outer boundary. 

\begin{figure}
\psfig{figure=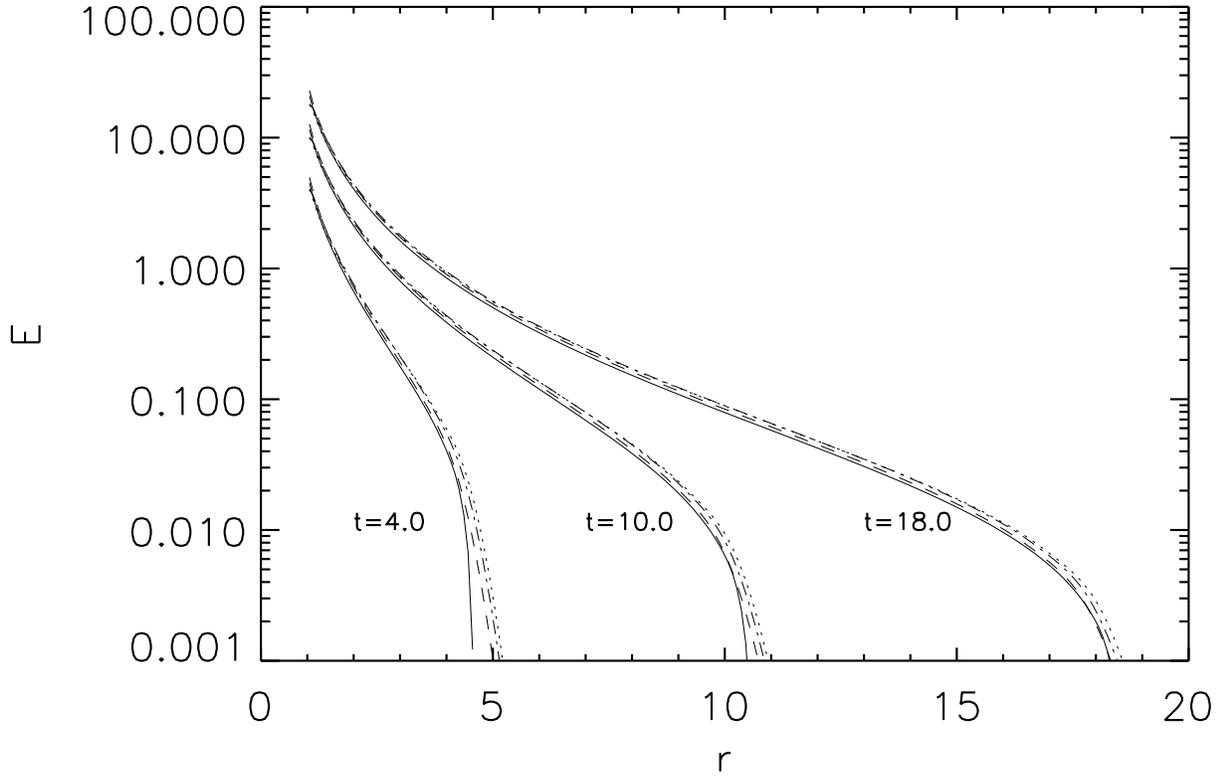,width=16cm}
\caption{Sphere radiating in the vacuum: Energy density as a function
of the radial coordinate for three different evolution times. The 
analytical solution (described in \S 5.3) is represented by solid lines
while the numerical solutions obtained with three different closures
are represented by dashes (vacuum), dots (Kershaw) and the dash-dotted
line (Minerbo). }
\label{fig3a}
\end{figure}

In Figure \ref{fig3a} the evolution of the radiation energy density as 
a function of the radial coordinate is shown for three different 
evolution times. The test has been repeat for three different
closure relations from those described in table 1: Vacuum (dashed lines),
Kershaw (dotted lines) and Minerbo (dashed-dotted lines). As in the 
previous figures the analytical solution is represented by the solid line.  
Although the appropriate closure for this problem (vacuum) gives a better
approximation to the real solution, it is remarkable that other closures
also give similar results. The differences between different closures
stem from the slightly different propagation speeds of the simple waves
in the regime of interest. Unlike in the two-wave problem described in 
previous section, for this specific problem any reasonable closure
might be used without major deviations from the real solution.

%%%%%%%%%%%%%%%%%%%%%%%%%%%%%%%%%%%%%%%%%%%%%%%%%%%%%%%%%%%%%%%%%%%%%%%%%%%%
\subsection{TEST 4: Radiative cooling and heating}

\begin{figure}
\psfig{figure=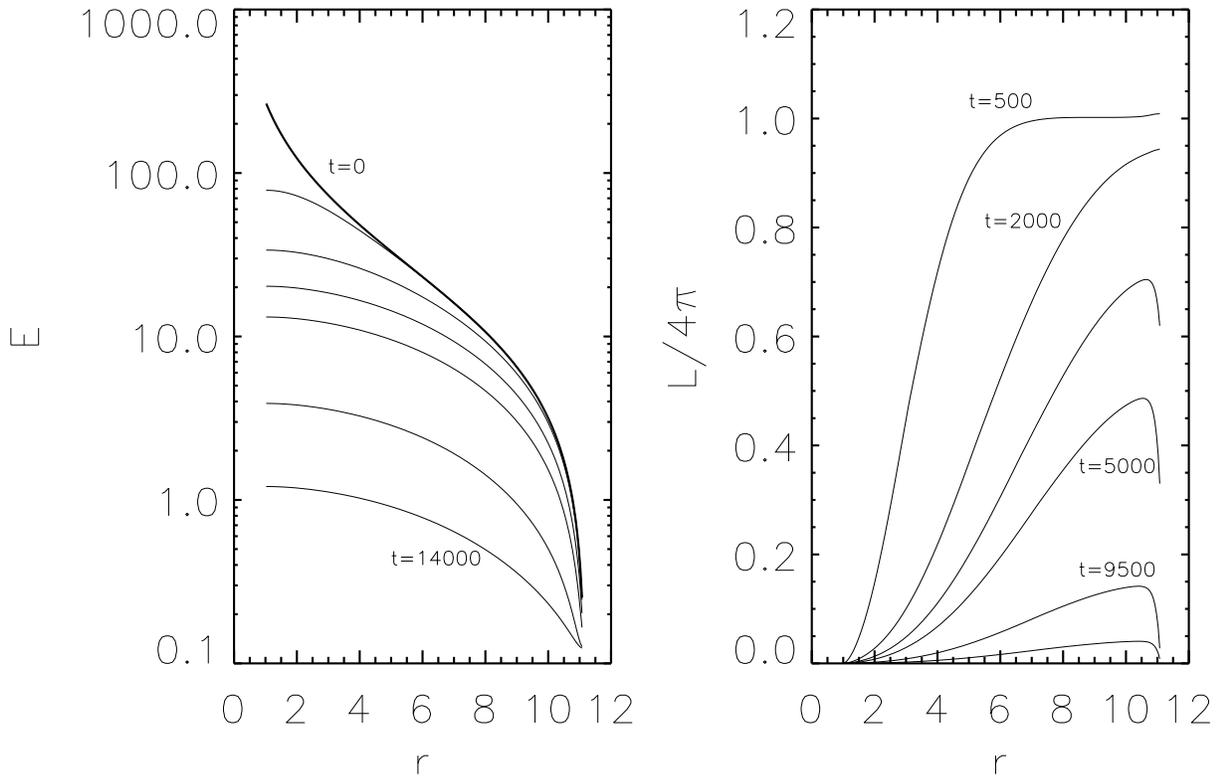,width=16cm}
\caption{Radiation initially in radiative equilibrium ($L=4 \pi$)
diffusing out after switching off the central energy source. The
solid thick line represents the initial model. The energy density
is plotted in the left panel and the luminosity in the right panel.}
\label{fig4a}
\end{figure}

\begin{figure}
\psfig{figure=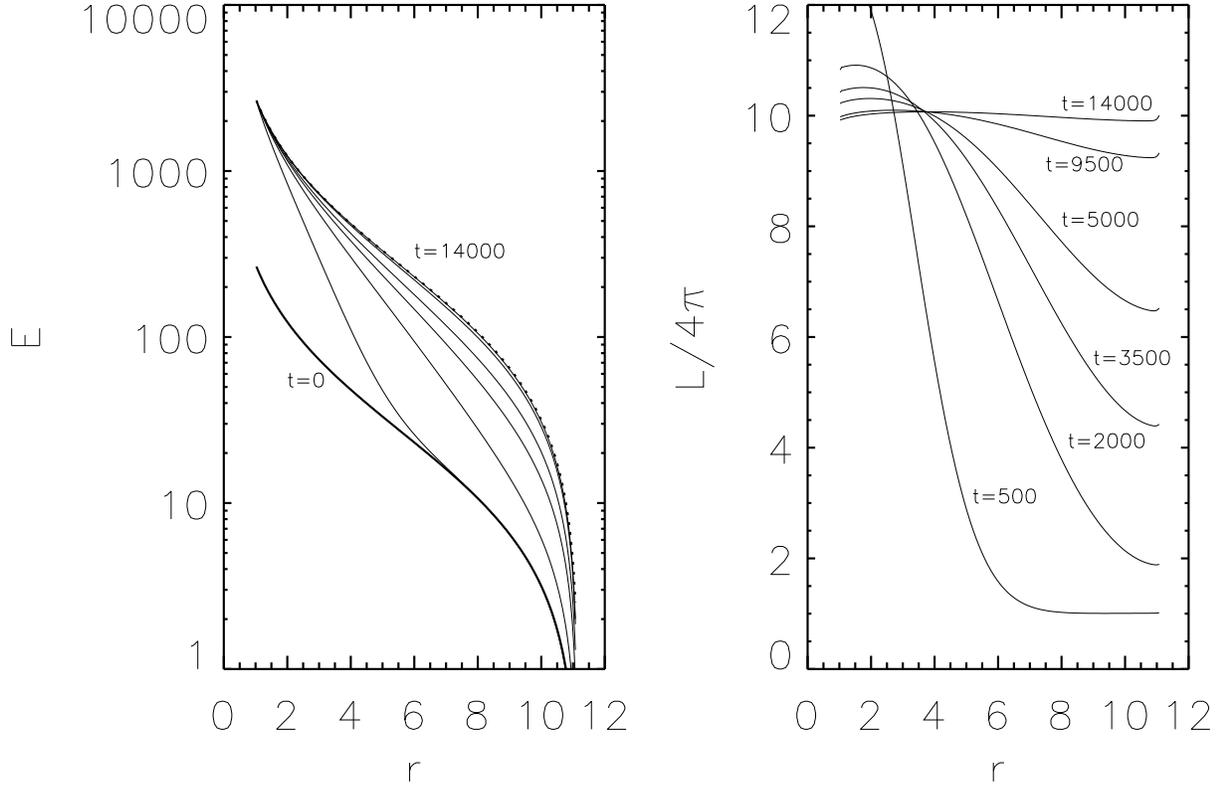,width=16cm}
\caption{Radiative heating: starting from the same initial model
(solid thick line) as in figure \ref{fig4a}, we increase the luminosity
at the inner boundary to $L=40 \pi$. The energy density evolves towards
the new stationary solution at higher luminosity.}
\label{fig4b}
\end{figure}

\begin{figure}
\psfig{figure=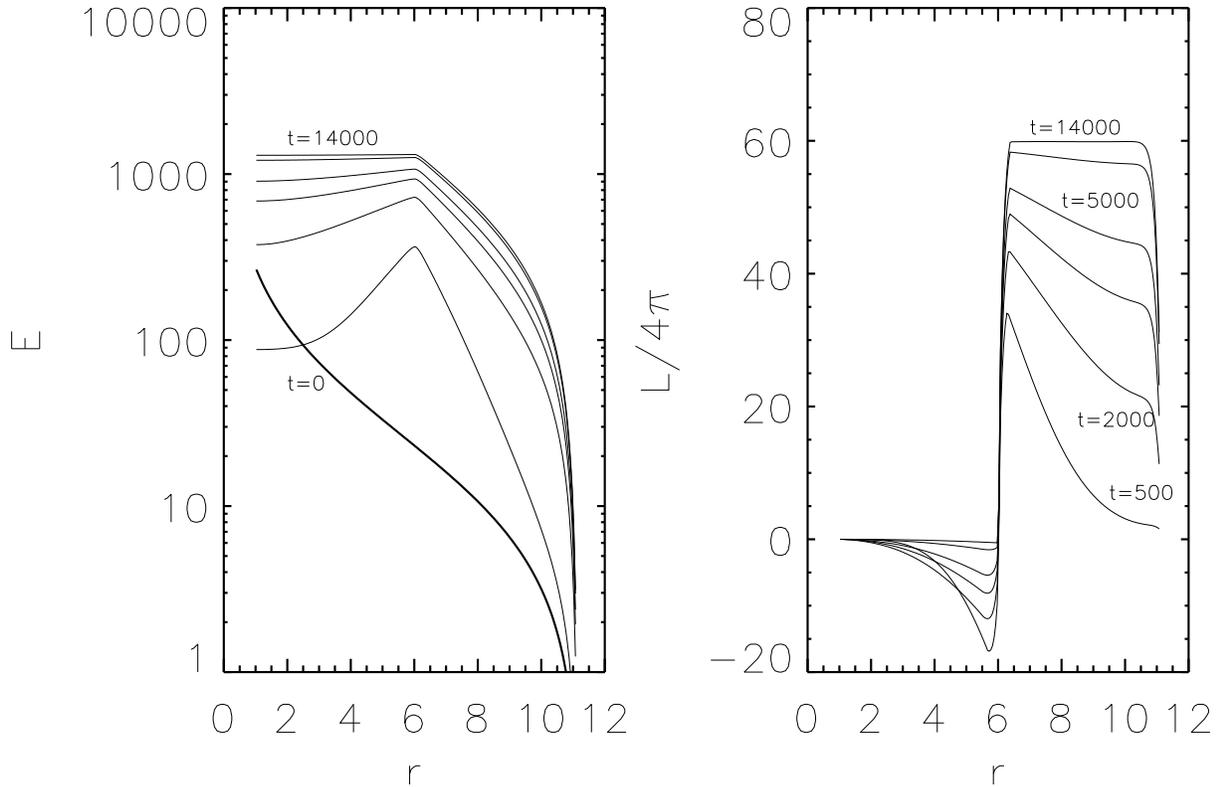,width=16cm}
\caption{Same as figure \ref{fig4a} but with a point--like energy source
located at $r=6$. The region inner to the source reaches the stationary
solution at $L=0$, while the outer region settles in radiative equilibrium
at luminosity higher than the initial model. The sharp discontinuity in the
luminosity is well resolved.}
\label{fig4c}
\end{figure}

With this test we intend to simulate a more realistic scenario in which,
initially, radiation produced in the center of a star diffuses out with
constant luminosity. We begin by
setting up a sphere which is in radiative equilibrium at a given
luminosity $L$. For $p=\frac{1}{3}$, the stationary solution is given by 
\be
E(r) = \frac{L}{4 \pi} \left[ \frac{1}{R^2} + 3 \kappa 
\left( \frac{1}{r}-\frac{1}{R} \right)  \right]
\ee
\be
F(r) = \frac{L}{4 \pi r^2}
\ee
where $R$ is the radius of the outer boundary, at which $f=1$.
Starting with the initial model corresponding to $L=4 \pi$, we
let it evolve in three different situations: 1) we take $L=0$ as boundary
conditions at the inner boundary (located at $r=1$), and the radiation field 
should diffuse out continuously; 
2) we change the inner boundary condition to
$L=40 \pi$, ten times larger than the initial model. The behaviour
is now the opposite, the energy density must increase until the
new stationary solution for the new luminosity is reached; 
3) we again take $L=0$ at the inner boundary but we switch on a 
central point--source in the middle of the star ($r=6$). 
The radiation field must evolve towards the corresponding stationary
solution, which is different in the two parts of the star defined by the
source. Inside, the luminosity is zero and the energy density is constant.
In the region external to the energy source, the stationary 
solution is defined by $L=\dot{s}$, where $\dot{s}$ is the
energy per unit time injected by the source.
The results for $\kappa=100$ are shown in Figures \ref{fig4a}-\ref{fig4c}, 
for cases 1 to 3, respectively. 
All three simulations have been done on the same grid of 200 uniform zones
and outflow boundary conditions in the outer edge have been imposed.

Left panels display the energy density as a function of radius.
The thick solid line in the figures corresponds to the initial model
and the thick dotted line in Figure \ref{fig4b} represent the stationary
solution with $L=40 \pi$. In Figure \ref{fig4b}, we can see how the
heat wave lasts about 1000 time-units (the diffusion time $R \kappa$)
to arrive to the surface and a
new stationary solution is successfully obtained after $\approx 15$ times
the diffusion time scale.

%%%%%%%%%%%%%%%%%%%%%%%%%%%%%%%%%%%%%%%%%%%%%%%%%%%%%%%%%%%%%%%%%%
\subsection{TEST 5: The semi-transparent regime.} 

Testing a method in the semi-transparent regime becomes a tough 
issue due to a lack of analytical solutions to compare with.
To overcome this problem, we have performed Monte Carlo simulations,
which is the closest solution to an exact one, and results in an
excellent test of the performance of the method in the
semi-transparent regime. 

We have used a Monte Carlo code that simulates the transfer of a
radiation field in a spherically symmetric geometry. We consider a 
region with inner boundary at $r = R_{in}$ and outer boundary at
$r = R_{out}$, divided in 200 spherical shells. Within 
each shell a constant value of the scattering opacity (elastic and isotropic)
is assumed. We do not consider emission or absorption processes. 
Our Monte Carlo procedure starts by injecting particles at the inner boundary 
outwardly-directed. Then, we compute the trajectory of each particle as 
it scatters off matter, according to the standard random walk laws 
\cite{Lucy99}. The 
trajectory ends when the particle escapes by crossing the outer boundary 
or it re-enters the inner boundary. Once the path of a particle is computed 
we proceed to calculate the contribution of this particle to the moments of 
the radiation field. The contribution of each particle to the energy density
at a given numerical shell is calculated by adding the path length between 
two successive scattering events, within the shell, divided by the volume 
of the shell. Consequently, the total energy density is obtained just by
adding the contribution of each particle multiplied by an arbitrary factor
Analogously, to obtain the contribution of each particle to the flux, 
the path length is weighted using the cosine of the angle with respect to the 
radial direction. Finally, pressure is calculated by means of the same 
procedure, but using the square of the cosine as weight of the path length.

In our numerical experiment we have taken: $R_{in} = 10$ and 
$R_{out} = 100$. The opacity law is an exponentially decreasing 
function in radius,
taking the value $\kappa=2$ at the inner boundary and $\kappa=2\times 10^{-5}$ 
at the outer boundary. 
The corresponding values of the optical depth are such that 
an important part of the region is semi-transparent. 
We have run the Monte Carlo code using $500\,000$ particles in order 
to obtain the stationary solution.

   The set up for our transport code is the following: 
i) the above opacity law, 
ii) the inner boundary conditions given by the stationary solution (obtained 
from Monte Carlo), and iii) the closure relation. We have performed 
two different calculations using Kershaw's 
closure relation and a cubic polynomial fit to Monte Carlo results, explicitly
\be
p = 0.3228+0.1902 f-0.0476 f^2 +0.5131 f^3
\ee
Starting with arbitrary initial profiles, we let 
the radiation field evolve until the stationary solution is reached and
results can be compared to those obtained from the Monte Carlo simulations.

In Figure \ref{fig6} we show the stationary solutions calculated from
Monte Carlo simulations (solid line) and from our code with the two
different closures, Kershaw's (dotted line) and the above fit (dashed line).
From the results displayed in the figure we can conclude that the
evolutionary code describes the semi-transparent regime accurately.
The differences in the values of the energy density (left panel),
normalized to its inner boundary value, amounts less than a few percentage in
any case. The slight discrepancies in the flux factor are mainly due to the
closure relation employed, being less than 6\% for Kershaw's closure and
less than 3\% for the fit.
The closure, therefore, sets the maximum accuracy that can be obtained, 
although qualitative results at a level of 10\% uncertainty can easily
be obtained with any reasonable closure in the semi-transparent regime.

%%%%%%%%%%%%%%%%%%%%%%%%%%%%%%%%%%%%%%%%%%%%%%%%%%%%%%%%%%%%%%%%%%
\subsection{A toy model.}
Our last experiment intends to illustrate how the numerical method
we proposed can succeed in situations when standard flux-limited
diffusion techniques fail.
We set up a sphere of radius $r=5$ with a radiation energy density of $E=1$,
embedded in a background where the opacity varies as follows: 
\be
\kappa = \left\{ {\matrix{
\kappa_0      &  r \le 1 \cr
\kappa_0/r^2  &  1 < r \le 10 \cr
0             &  10 < r \le 15 \cr
\kappa_0      &  15 < r \le 17 \cr
0             &  r \ge 17 \cr
}} \right.
\ee
where we have taken $\kappa_0=10^3$. Outside the sphere we take $E=0$
(in practice, $E=10^{-6}$ for numerical reasons), 
and initially the flux is zero everywhere.
It tries to mimic a situation in which radiation scatters out in
a central star, where the diffusion approximation remains valid, through
an opacity-varying atmosphere up to the surface ($r=10$), where the
opacity suddenly vanishes. Then, it crosses a transparent region to
find a very opaque layer, where diffusion operates again.
This sort of problem is obviously a tough challenge for diffusion-based
codes, since they cannot work properly in the transparent region. 

In Figure \ref{fig6} we show results from the evolution of this problem
on a grid of 200 uniform shells and using Kershaw's closure. In the left
panel we show the energy density for different times. 
It can be seen how a gradient is developed in the diffusive regions
$(r<10, 15<r<17)$ while radiative equilibrium $(dE/dr=0)$ in the transparent
region between the two opaque layers is rapidly obtained.
Notice also that only for the last snapshot ($t=1000$, dashed curve) there is
a flux of energy in the outer transparent region, because it takes some
time to the radiation to diffuse through the opaque layer and develop
a gradient that keeps the outgoing flow. 

%%%%%%%%%%%%%%%%%%%%%%%%%%%%%%%%%%%%%%%%%%%%%%%%%%%%%%%%%%%%%%%%%%

\section{Conclusions}

We have studied the mathematical character of the angular moment equations of 
radiative transfer, considering the implications of
using different closure relations.
The role played by the closure and their derivatives is more apparent
in the hyperbolic formulation.
After this analysis, we conclude that the hyperbolic character is assured for 
a given variety of closures widely used in the literature: those given by
$p=p(f)$. For general closures $p=p(e,f)$, hyperbolicity can not be guaranteed,
although the closure proposed by Cernohorsky \& Bludman \shortcite{CB94},
based on maximum entropy arguments, leads to real eigenvalues.
Additional constraints on the closures are imposed on the base of the 
behaviour of the eigenvalues of the Jacobian matrix since they give the 
velocity of propagation of perturbations. Causality limitation is written in 
terms of the closure relation helping us to select, from this point of view,
among the different closures reviewed. It turns 
out that only one of the closures analysed does not satisfy causality,
allowing for velocities of the waves higher than the speed of light. 
For the other closures, 
although derived without taking into account
the mathematical issues addressed in this paper,
we found that they are consistent with hyperbolicity and causality.

Writing the moment equations of the distribution
function as a hyperbolic system of conservation laws
allows one to apply numerical techniques specifically
designed for such systems, the so-called HRSC methods.
We have applied HRSC methods to solve the transport equations in a static 
background showing, for a number of test problems, the feasibility of the
method. From the results, the main conclusion is that the numerical method
can be used in any regime, from optically thick to transparent regions,
obtaining numerical solutions with 
high accuracy. It turns out that the closure relation plays an important 
role, more evident than in traditional methods for radiation
transfer, that can be useful to choose the more appropriate closure for 
a given problem.

When the radiation transport equations are written as a system of conservation
laws, the coupling with hydrodynamics is straightforward 
\cite{Morel99}. This feature make us 
to be confident in the possibility of applying the method to problems 
involving radiative flows. Moreover, 
we are optimistic in the applicability to multidimensional problems,
since the extension of the method is straightforward.

%%%%%%%%%%%%%%%%%%%%%%%%%%%%%%%%%%%%%%%%%%%%%%%%%%%%%%%%%%%%%%%%%%%%%%%%%%%%%
\section*{Acknowledgements}
{This work has been supported by the Spanish
DGICYT (grant PB97-1432). J.A.P. also acknowledges previous financial 
support from the Ministerio de Educaci\'on y Cultura.}

%%%%%%%%%%%%%%%%%%%%%%%%%%%%%%%%%%%%%%%%%%%%%%%%%%%%%%%%%%%%%%%%%%%%%%%%%%%%%

\newpage

\begin{center}
\begin{table}
\caption[]{Closure and characteristic structure.}
\begin{tabular}{|c|cccccc|}
\hline \hline
\, & $p(f=0)$ & $p(f=1)$ & $p'(f=0)$ & $p'(f=1)$ & $\lambda_{\pm} (f=0) $
& $\lambda_{\pm} (f=1) $
\\
\hline
Minerbo & ${1/3}$ & 1 & 0 & 2 & $(\frac{1}{\sqrt{3}}, -\frac{1}{\sqrt{3}})$
		& (1,1) \\
LP      & ${1/3}$ & 1 & 0 & 1 & $(\frac{1}{\sqrt{3}}, -\frac{1}{\sqrt{3}})$
		& (1, 0) \\
Auer    & ${1/3}$ & 1 & ${2/3}$ & ${2/3}$ & $(1, -\frac{1}{3})$
		& $(1, -\frac{1}{3})$ \\
Kershaw & ${1/3}$ & 1 & 0 & ${4/3}$  &$(\frac{1}{\sqrt{3}}, -\frac{1}{\sqrt{3}})$
		& $(1, \frac{1}{3})$ \\
Bruenn  & ${1/3}$ & 1 & $-{1/3}$ & ${5/3}$ & $(-1\pm \sqrt{13})/6$
		& $(1, \frac{2}{3})$ \\
BW      & ${1/3}$ & 1 & $-{4/3}$ & ${5/3}$ & $(-2\pm \sqrt{7})/3$ 
		& $(1, \frac{2}{3})$ \\
Janka   & ${1/3}$ & 1 & 0 & 2 & $(\frac{1}{\sqrt{3}}, -\frac{1}{\sqrt{3}})$
		& (1,1) \\
Vacuum  & ${1/3}$ & 1   & $-{2/3}$ & 2 & $(-1, \frac{1}{3})$	& (1,1) \\
\hline \hline
\end{tabular}
\end{table}
\end{center}

\newpage

\begin{figure}
\psfig{figure=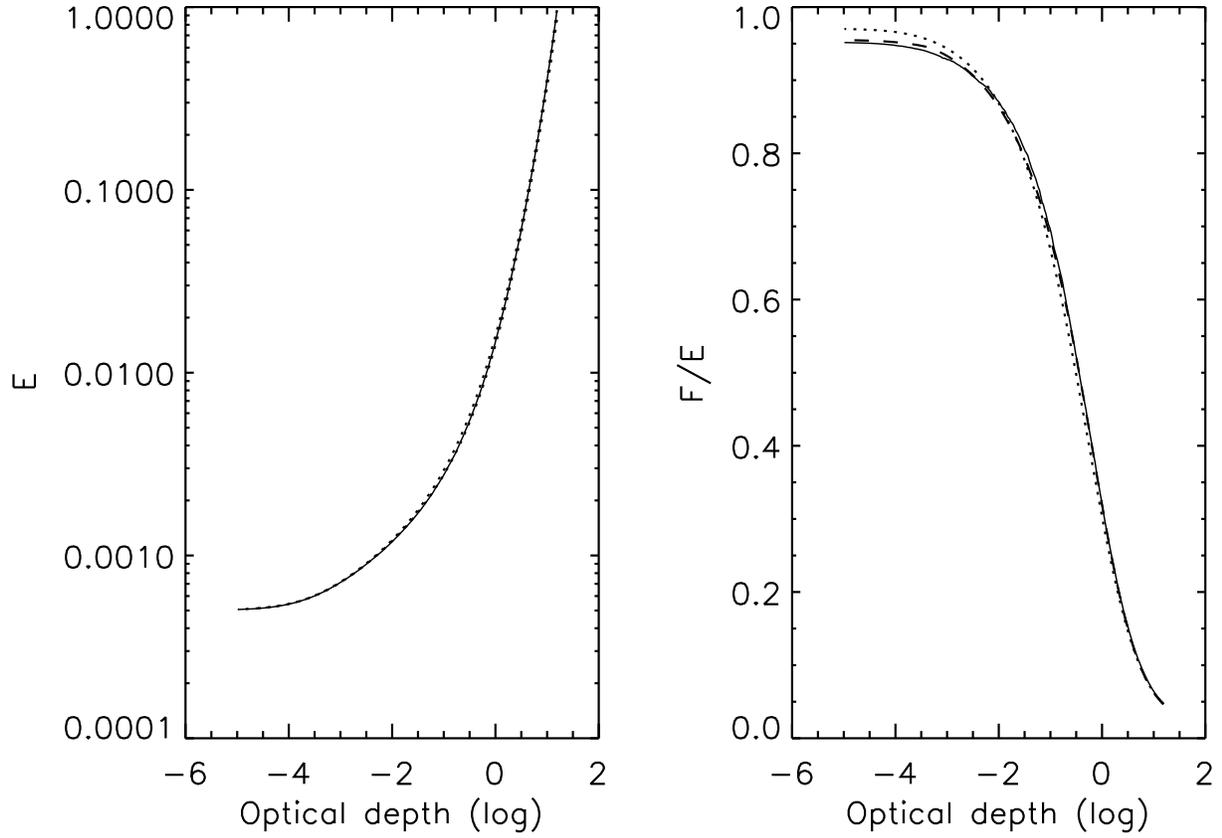,width=16cm}
\caption{
Semi-transparent regime: Stationary solution of 
the radiation field in a background
with a scattering opacity exponentially decreasing with radius. 
Solid lines correspond to the Monte Carlo simulation and dotted and
dashed lines correspond to the results obtained with the 
evolutionary code with Kershaw's closure and a numerical fit to
Monte Carlo results, respectively. Left panel shows
the normalized (to its inner boundary value) energy
density as a function of the optical depth.
Right panel shows the flux factor. 
}
\label{fig5}
\label{fnueva}
\end{figure}

\begin{figure}
\psfig{figure=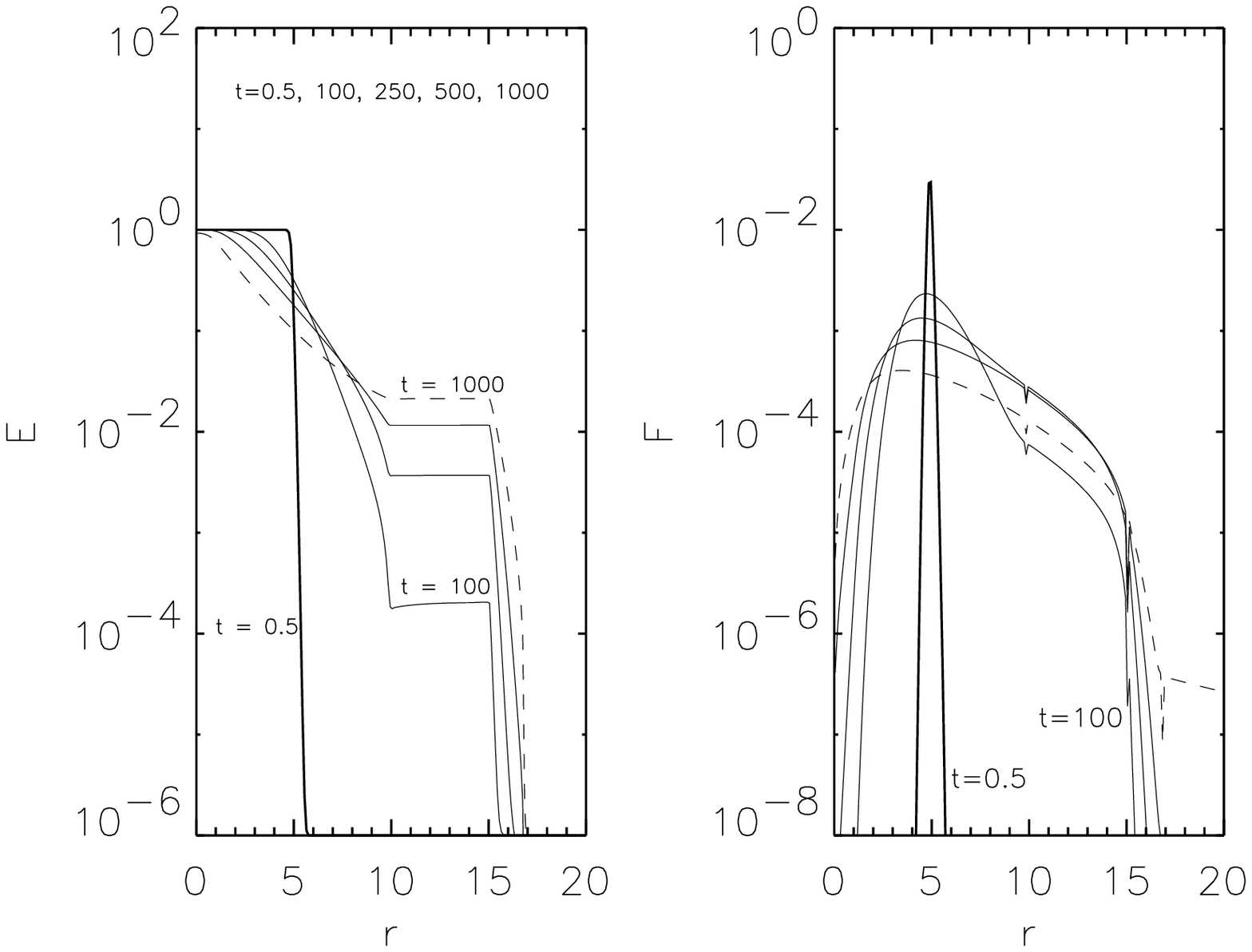,width=16cm}
\caption{Evolution of the radiation field in a background
with strongly varying opacity. Snapshots of the energy density (left panel)
and flux (right panel) are shown for different times 
$t=0.5, 100, 250, 500, 1000$. The thick curve corresponds to $t=0.5$
and the dashed curve to $t=1000$. See the text for
details on the opacity profile.}
\label{fig6}
\end{figure}


\begin{thebibliography}{}

\bibitem[\protect\citename{Auer }1984]{Auer84}
Auer, L. H., 1984, in Methods in Radiative Transfer, 
ed. W. Kalkofen, Cambridge University Press, Cambridge.

\bibitem[\protect\citename{Banyuls {\it et al.} }1997]{BFIMM97}
Banyuls, F., Font, J. A., Ib\'a\~{n}ez J.M$^{\underline{\mbox{a}}}$., 
Mart\'{\i}, J. M$^{\underline{\mbox{a}}}$., Miralles, J. A., 
1997, ApJ, 476, 221.

\bibitem[\protect\citename{Blinnikov \& Bartunov }1993]{BB93}   %JOSE
Blinnikov, S. I., Bartunov, O. S., 1993, A\&A., 273, 106

\bibitem[\protect\citename{Bowers \& Wilson }1982]{BW82}
Bowers, R. L., Wilson, J. R., 1982, ApJ Suppl., 50, 115

\bibitem[\protect\citename{Bruenn }1985]{Bru85}
Bruenn S. W., 1985, ApJ Suppl., 58, 771

\bibitem[\protect\citename{Cernohorsky \& Bludman }1994]{CB94}
Cernohorsky, J., Bludman, S. A., 1994, ApJ, 433, 250

\bibitem[\protect\citename{Cooperstein, van den Horn \& Baron }1986]{CVB86}
Cooperstein, J., van den Horn, L.J., Baron, E.A., 
1986, ApJ, 309, 653.

\bibitem[\protect\citename{Dgani \& Janka }1992]{DJ92}
Dgani, R., Janka, H.-Th., 1992, A\&A, 256, 428

\bibitem[\protect\citename{Eastman \& Pinto }1993]{EP93}   %JOSE
Eastman, R. G., Pinto, P. A., 1993, ApJ, 412, 731

\bibitem[\protect\citename{Ensman \& Burrows }1992]{EB92}   %JOSE
Ensman, L., Burows, A., 1992, ApJ, 393, 742

\bibitem[\protect\citename{Gehmyer \& Mihalas }1994]{GM94}   %JOSE
Gehmeyr, M., Mihalas, D., 1994, Physica D, 72, 320

\bibitem[\protect\citename{Godunov }1959]{Go59}
Godunov, S. K., 1959, Mat. Sb., 47, 271.

\bibitem[\protect\citename{Groth {\it et al.} }1996]{Gro96}
Groth, C. P. T., Gombosi, T. I., Roe, P., L., Brown, S. L., 1996,
Proceedings of the {\it Fifth international conference on
Hyperbolic Problems}, Eds. Glimm, J., Graham, M.J., Grove, J.W.,
World Scientific.

\bibitem[\protect\citename{Harten, Lax \& van Leer }1983]{HLL83}
Harten, A., Lax, P.D., van Leer, B., 1983, SIAM Rev., 25, 35.

\bibitem[\protect\citename{Ib\'a\~{n}ez \& Mart\'{\i} }1999]{IM99}
Ib\'a\~{n}ez, J. M$^{\underline{\mbox{a}}}$., 
Mart\'{\i}, J. M$^{\underline{\mbox{a}}}$., 1991, 
J. Comput. Appl. Math., in press.

\bibitem[\protect\citename{Janka }1991]{Jan91}
Janka H.-Th., 1991, PhD thesis, Technische Universit\"at M\"unchen.

\bibitem[\protect\citename{Janka, Dgani \& van den Horn }1992]{JDH92}
Janka H.-Th., Dgani, R., van den Horn, L. J., 1992, A\&A, 265, 345.

\bibitem[\protect\citename{Jin \& Levermore }1996]{JL96}
Jin, S., Levermore, C.D., 1996, J. Comput. Phys., 126, 449.

\bibitem[\protect\citename{Kershaw }1976]{Ke76}
Kershaw, D. S., 1976, report UCRL-78378, Lawrence Livermore
National Laboratory, Livermore, CA.

\bibitem[\protect\citename{Kunasz  }1983]{Ku83}
Kunasz, P. B., 1983, ApJ, 271, 321.

\bibitem[\protect\citename{van Leer }1979]{vL79}
van Leer, B., 1979, J. Comput. Phys., 32, 101.

\bibitem[\protect\citename{LeVeque }1992]{Le92}
LeVeque R. J., 1992, 
Numerical methods for conservation laws, Birkh\"auser, Basel.

\bibitem[\protect\citename{Levermore \& Pomraining }1981]{LP81}
Levermore, C.D., Pomraining, G. C., 1981, ApJ, 248, 321

\bibitem[\protect\citename{Lindquist }1966]{Lin66}
Lindquist, R. W., 1966, Ann. Phys., 37, 478

\bibitem[\protect\citename{Lowrie, Morel \& Hittinger }1999]{Morel99}
Lowrie, R.B., Morel, J.E., Hittinger, J.A., 1999, ApJ, 432, 521.

\bibitem[\protect\citename{Lucy }1999]{Lucy99}
Lucy, L.B., 1999, A\&A, 344, 282.

\bibitem[\protect\citename{Mart\'{\i}, Ib\'a\~{n}ez \& Miralles }1991]{MIM91}
Mart\'{\i}, J. M$^{\underline{\mbox{a}}}$., Ib\'a\~{n}ez, 
J. M$^{\underline{\mbox{a}}}$., Miralles, J. A., 1991, Phys. Rev., D43, 3794

\bibitem[\protect\citename{Mart\'{\i} \& M\"uller }1999]{MM99}
Mart\'{\i}, J. M$^{\underline{\mbox{a}}}$.,  
M\"uller, E., 1999, {\it to appear in} Living Reviews in Relativity,
Vol. 2.

\bibitem[\protect\citename{Mezzacappa \& Bruenn }1993]{MB93}
Mezzacappa, A., Bruenn, S. W., 1993, ApJ, 405, 669

\bibitem[\protect\citename{Mihalas \& Kunasz }1986]{MK86}
Mihalas, D., Kunasz, P. B., 1986, JCP, 64, 1M.

\bibitem[\protect\citename{Mihalas \& Mihalas }1984]{MM84}
Mihalas, D., Mihalas, B. W., 1984, Foundations of Radiation
Hydrodynamics, Oxford Univ. Press, New York.

\bibitem[\protect\citename{Mihalas {\it et al.} }1976]{MSKH76}
Mihalas, D., Shine, R.A., Kunasz, P. B., Hummer, D. G., 1976, ApJ, 205, 492.

\bibitem[\protect\citename{Minerbo }1978]{Mi78}
Minerbo, G. N., 1978, J. Quant. Spectrosc. Radiat. Transfer, 20, 541

\bibitem[\protect\citename{Miralles, Van Riper \& Lattimer }1993]{MRL93}
Miralles, J. A., Van Riper, K. A., Lattimer, J. M., 1993, ApJ, 407, 687.

\bibitem[\protect\citename{Norman \& Winkler }1986]{NW86}
Norman, M. L., Winkler, K-H. A., 1986, 
Astrophysical Radiation Hydrodynamics, Reidel.

\bibitem[\protect\citename{Pons {\it et al.} }1998]{PFIMM98}
Pons, J. A., Font, J. A., Ib\'a\~{n}ez J.M$^{\underline{\mbox{a}}}$., 
Mart\'{\i}, J. M$^{\underline{\mbox{a}}}$., Miralles, J. A., 
1998, A\&A, 339, 638.

\bibitem[\protect\citename{Schinder \& Bludman }1989]{ScB89}
Schinder, P. J., Bludman S. A., 1989, ApJ, 346, 350

\bibitem[\protect\citename{Thorne }1981]{Tho81}
Thorne K. S., 1981, MNRAS, 194, 439

\bibitem[\protect\citename{Turolla \& Nobili }1988]{TN88}
Turolla, R., Nobili, L., 1988, MNRAS, 235, 1273

\bibitem[\protect\citename{Turolla, Zampieri \& Nobili }1995]{TZN95}
Turolla, R., Zampieri, L., Nobili, L., 1995, MNRAS, 272, 625

\bibitem[\protect\citename{Yamada, Janka \& Suzuki }1999]{Ya99}
Yamada, S., Janka, H.-Th., Suzuki, H., 1999, A\&A, 344, 533.

\end{thebibliography}
\end{document}